\documentclass[11pt,showpacs,showkeys,preprintnumbers]{revtex4-2}

\usepackage{graphicx}   
\usepackage{color}      
\usepackage{braket}
\usepackage{amsmath}
\usepackage{amssymb}
\usepackage{mathrsfs} 
\usepackage{titlesec}
\usepackage{makecell}
\usepackage{natbib,hyperref}
\usepackage[mathscr]{euscript}

\def \be  {\begin{equation}}
\def \ee  {\end{equation}}
\def \ee  {\end{equation}}
\def \bea {\begin{eqnarray}}
\def \eea {\end{eqnarray}}

\newcommand{\R}{\mathbb{R}}
\DeclareMathOperator{\sech}{sech}
\usepackage{mathrsfs}  

\definecolor{ao-english}{rgb}{0.0, 0.5, 0.0}
\definecolor{cadmiumgreen}{rgb}{0.0, 0.42, 0.24}

\newcommand{\nn}{\nonumber}

\begin{document}

\preprint{ECTP-2024-05}
\preprint{WLCAPP-2024-05}
\preprint{FUE-2024-05}
\hspace{0.05cm}

\title{Expansion Evolution of Nonhomogeneous Metric with Quantum-Mechanically Revisited Fundamental Metric Tensor}

\author{Abdel Nasser Tawfik $^{1,\ast}$, Azzah A. Alshehri $^{2}$, and Antonio Pasqua$^{2}$}
\email{a.tawfik@fue.edu.eg}
\affiliation{%
$^{1}$ Basic Science Department, Faculty of Enginering, Future University in Egypt (FUE), 11835 New Cairo, Egypt \\ 
$^{2}$ Physics Department, University of Hafr Al Batin, Hafr Al Batin 39524, KSA\\
$^{3}$ Department of Physics, University of Trieste, 34100 Trieste, Italy. }

\begin{abstract}
To explore the properties of space and initial singularities in the context of general relativity, where spacetime becomes poorly defined and no longer belongs to a regular manifold, we examine the evolution of the expansion of timelike geodesic congruences for two distinct formulations of the fundamental metric tensor. This analysis is conducted within a nonhomogeneous, anisotropic, and spherically symmetric cosmic background. The results derived from the conventional metric tensor, the building block of Einstein's theory of general relativity, are compared with those obtained from a quantum-mechanically revisited metric tensor. This comparison enables an assessment of the proposed geometric quantization, particularly in terms of whether singularities are regulated or diminished. Utilizing a quantum geometric approach, the numerical analysis incorporating a quantum-mechanically revisited metric tensor applies a mean-field approximation on the integrated quantum operators. In contrast to the results obtained with conventional metric tensor, the quantum-mechanically induced revision of the metric tensor seems to provide a framework for controlling singularities in the new formulation of general relativity. The degree of quantization likely influences the ability to regulate or even potentially remove both singularities. We also conclude that the proposed geometric quantization provides a means to explore the quantum nature of spacetime curvatures, emphasizing that the singularity dilemma arose primarily from the standard semi-classical approximation of Einstein's general relativity.

\end{abstract}
\keywords{Einstein--Gilbert--Straus Metric; Space and initial singularities;  Discretized curved spacetime; Riemann--Finsler--Hamilton geometry}

\maketitle


\section{Introduction}
\label{sec:intro}

Over the course of several decades, numerous attempts have been undertaken to merge two distinct solutions of the Einstein Field Equation (EFE). These solutions include spherically symmetric, homogeneous, isotropic metric (Schwarzschild) and the Friedmann--Lemaitre--Robertson--Walker (FLRW) metric \cite{Tawfik:2017ngn}, as explored by McVittie \cite{McVittie:1933zz}, Tolman \cite{Tolman1934}, Einstein and Strauss \cite{RevModPhys.17.120}, Bondi \cite{Bondi:1947fta}, and Gilbert \cite{Gilbert1956}. It was Gilbert who ultimately proposed a successful formalism known as the Einstein--Gildert--Straus (EGS) metric \cite{Gilbert1956,Stewart:1990ngn}. This metric combines the Schwarzschild metric, which represents the most general spherically symmetric vacuum solution of EFE and describes the gravitational field surrounding a spherical mass distribution, with the FLRW metric, which characterizes an expanding Universe that is both homogeneous and isotropic.

The EGS metric provides a description of the Universe as clusters of cosmic substance interspersed with voids and holes. The cosmic substance is contained within the voids, with the quantity of substance in the centers of the voids being equal to the amount excavated to form the voids, resulting in an nonhomogeneous Universe. This metric extends the standard cosmological model by allowing for nonhomogeneity in the expanding Universe. The evolution of the nonhomogeneous Universe is contingent on both radial distance $r$ and cosmic time $t$. The timelike geodesic congruence is similarly reliant on both $r$ and $t$. To explore the interdependence of $r$ and $t$, the present calculations assume that the cosmic substance within the voids can be modeled as forming perfect spheres, with radii that vary with cosmic time. The assumption regarding the latter is contingent upon the specific model being utilized. The present study aims to investigate the evolution of expansion utilizing timelike geodesic congruence and compare the results obtained from the conventional fundamental tensor $g_{\alpha\beta}$ with those obtained from the quantum-mechanically revisited fundamental metric tensor $\tilde{g}_{\alpha\beta}$.

The theory of generalized geometric structures, proposed by of of the authors (AT), has been examined in various cosmological phenomena \cite{NasserTawfik:2024afw,Tawfik:2023onh,Tawfik:2023ugm,Tawfik:2023hdi}. This theory involves multiple concepts, including duality-symmetry configurations, Finsler--Hamilton manifold, noncommutative algebra, and quantum geometry within the framework of general relativity (GR) \cite{Tawfik:2025kae,Tawfik:2024itt,Tawfik:2023ugm,Tawfik:2023hdi}. Specifically, the four-dimensional Riemann manifold was extended to the phase-space structured Finsler--Hamilton manifold to incorporate quantum-mechanical elements. This expansion allowed for the inclusion of auxiliary four-vectors of coordinates and momenta for free-falling quantum particles, denoted as $x_0^{\alpha}$ and $p_0^{\beta}$, respectively \cite{Tawfik:2023kxq,Tawfik:2023rrm,Farouk:2023hmi,Tawfik:2021ekh}. By incorporating the relativistic generalized uncertainty principle (RGUP) to address the effects of relativistic energy and gravitational field on quantum mechanics (QM), the momentum operator $p_0^{\nu}$ was generalized to $\phi(p) p_0^{\nu}$, where $\phi(p)=1+\beta p_0^{\rho} p_{0 \rho}$ \cite{Tawfik:2024gxc}. By examining the Hessian matrix of the squared Finsler (Hamilton) structure, denoted as $F^2(x_0^{\alpha}, \phi(p) p_0^{\beta})$, the corresponding metric on a Finsler manifold can be obtained. Conversely, on a Riemann manifold, the quantum mechanical revision of metric tensor can be found by equating the line elements of both manifolds. $\rho$ is a dummy index. $\alpha, \beta \in \{0,1,2,3\}$ are free indexes \cite{Tawfik:2023rrm,Tawfik:2023kxq}. By conducting table-top experiments and analyzing astronomical observations, the RGUP parameter, $\beta$ (not the index), and thereby the degree of quantization can be determined.

The spacetime curvature is characterized by the evolution of a family of trajectories, as indicated in ref. \cite{Kolomytsev:1972rf}. These trajectories form a trajectory congruence, which is represented by flow lines generated by velocity fields \cite{Tsamparlis:1995nq}. These flow lines correspond to the flows of world lines generated by vector fields, which can be either geodesic or nongeodesic. The primary goal of this research is to characterize the timelike geodesic congruence using the EGS metric, incorporating the conventional metric tensor $g_{\alpha \beta}$ and the quantum mechanical revision of metric tensor $\tilde{g}_{\alpha \beta}$. The process of utilizing the evolution of the EGS metric expansion is notably challenging, mainly owing to its complicated and non-homogeneous structure. The complex interplay between the radial distance $r$ and cosmic time $t$ adds another layer of complexity to the challenge \cite{Tawfik:2023onh}.

It is well known that the presence of spacetime singularities is a common phenomenon within the semi-classical approximation of general relativity \cite{Hawking:1970zqf}. This situation gives rise to a form of geodesic incompleteness that is observed in various classes of solutions of EFE \cite{Landsman:2022hrn}. Attempts to quantize GR using conventional quantum field theory have highlighted the perturbative non-renormalizability of the theory \cite{Borissova:2023kzq}. Nonetheless, it is in the context of quantum gravity that one might expect to uncover their essential nature \cite{DeWitt:1967yk}. The study of how spacetime quantization influences singularities can be approached through techniques derived from (2+1)-dimensional quantum gravity \cite{Nelson:1991an}. The present script explores the characteristics of both space-like and timelike singularities based on a geometric quantization and shows how the geometric quantization approach is able to attenuate or regulate the singularities in GR.

This manuscript is structured as follows. Section \ref{sec:frmlsm0} provides an outline of the formalism. The quantum-mechanically induced revision of fundamental metric tensor $\tilde{g}_{\alpha\beta}$ shall be introduced in section \ref{sec:tildegmunuApp}. 
The derivation of the quantum mechanical revision of Finsler--Hamilton metric is discussed in section \ref{sec:FnslMtr}. In section \ref{sec:EGS1}, the Einstein-Gilbert-Straus (EGS) metric is introduced. The derivation of the timelike geodesic congruence and expansion evolution using conventional $g_{\alpha \beta}$ and quantum mechanically revisited metric tensor $\tilde{g}_{\alpha \beta}$ can be found in section \ref{sec:egsSol1} and \ref{sec:egsSol2}, respectively. Some aspects of the quantum mechanical revision and mean-field approximation are elaborated in section \ref{sec:qMFA}. The discussion of the numerical results is presented in section \ref{sec:dsct}. Finally, section \ref{sec:cncl} concludes this study. 
In Appendix \ref{sec:App1}, we discuss the geodesic equations and their fulfillment by the metric $\tilde{g}_{\alpha\beta}$. By establishing that $\tilde{g}_{\alpha\beta}$ complies with the geodesic equations, it becomes evident that EFE are consequently satisfied.

\section{Formalism}
\label{sec:frmlsm0}

\subsection{Quantum-Mechanical Revision of Metric Tensor}
\label{sec:tildegmunuApp}

The relativistic generalized uncertainty principle (RGUP) is a crucial element in the generalization of quantum mechanics, as it integrates the effects of gravity and introduces a novel form of uncertainty that operates within the realm of four-dimensional spacetime \cite{Tawfik:2024gxc}. In RGUP, the momentum $p_0^{\nu}$ undergoes deformation to $\phi(p) p_0^{\nu}$, where $\phi(p) = 1 + 2 \beta p_{0}^{\rho}p_{0 \rho}$, with $\beta$ representing the RGUP parameter and $p_0$ denoting the auxiliary four-vector of momentum for a free-falling quantum particle with positive mass $m$. Additionally, in order to incorporate the kinematics of the free-falling quantum particle based on its positive homogeneity, the Finsler (Hamilton) structure can also be extended. This extended Finsler (Hamilton) structure becomes discretized and is characterized by coordinates $x_0^{\alpha}$ and direction coupled to the RGUP-deformed momentum of the free-falling quantum particle, $\phi(p) p_0^{\beta}$. Through a quantum geometrical approach, the four-dimensional fundamental tensor on a Riemannian manifold can be derived from the eight-dimensional Finslerian metric, which includes quantum-mechanically induced modifications 
\bea
\tilde{g}_{\mu \nu}  &=& \left(\phi^2(p)+2\frac{\kappa}{(p_0^0)^2} K^2\right) 
 \left[1 + \frac{\dot{p}_0^{\mu} \dot{p}_0^{\nu}}{{\mathscr F}^2} \left(1+2\beta p_0 ^{\rho} p_{0 \rho}\right)  \right] g_{\mu\nu} \nn \\
&+& \left[\frac{d x_0^{\mu}}{d \zeta^{\mu}} \frac{d x_0^{\nu}}{d \zeta^{\nu}} +  \left(1+2\beta p_0 ^{\rho} p_{0 \rho}\right) \frac{d p_0^{\mu}}{d \zeta^{\mu}} \frac{d p_0^{\nu}}{d \zeta^{\nu}} \right] d_{\mu\nu}(x,p).  \label{eq:gmunuQ2}
\eea
Details about the derivation of $\tilde{g}_{\mu \nu}$ can be found in refs. \cite{Tawfik:2023hdi,Tawfik:2023ugm,Tawfik:2023rrm,Tawfik:2023kxq}. The concept of {\it ''geometry''} is associated with the dependence on $\dot{p}_0^{\mu} \dot{p}_0^{\nu}$, or what is termed {\it ''acceleration''}, where $\dot{p}_0$ represents a gravitational force acting on the free-falling quantum particle in curved spacetime. This quantity is not invariant under the reparametrization of the curve parameter, thereby resulting in a pathological parametrization. The existence of quantum spacetime is a consequence of the generalized uncertainty principle.
Analogous to the indexes $\alpha$ and $\beta$, the indexes $\mu, \nu \in \{0,1,2,3\}$ are treated as free indexes. The variable $\zeta^{\mu}$ serves as a parametrization that establishes a link between the coordinates in the Finslerian tangent bundle and the Riemann coordinates. It is of utmost importance to highlight that in order to ensure the equivalence of the Finslerian and Riemannian measures of the line element should be harmonized. Furthermore, the expression for $d_{\mu\nu}(x,p)$ in the second line of Eq. \eqref{eq:gmunuQ2}
\bea
d_{\mu\nu}(x,p) &=&  \frac{6 \kappa \phi(p)}{(p_0^0)^2}\left\{K^2 \ell_{\mu} \ell^{\sigma} g_{\sigma \mu} - K^3 \ell^{\sigma} \left[\delta_{0 \nu} g_{\sigma\mu} + \delta_{0 \mu} g_{\sigma\nu}\right] 
+ \frac{2+\phi(p)}{8 \phi(p)} K^4 \delta_{0 \mu} \delta_0^{\sigma} g_{\sigma\nu}\right\}, 
\eea 
needs to be redefined in terms of $g_{\mu\nu}(x,p)$. This redefinition is necessary to maintain consistency with the first line of Eq. \eqref{eq:gmunuQ2}, where the line measures are already identical. Until the resolution of this mathematical challenge, it is necessary to make a bold estimation that the second line diminishes, at some scales. However, it is important to note that the function $\phi(p)$ is independent on both $x_0$ and $d_{\mu\nu}$. This means that $\phi(p)$ remains finite, which means that $\phi(p)$ assumes certain limitations on such a bold approximation. In order to thoroughly examine the potential consequences of the proposed geometric quantization approach in proving or disproving the existence of space and initial singularity, and in the absence of any viable alternative to keeping $\tilde{g}_{\mu \nu}$, Eq. \eqref{eq:gmunuQ2}, non-truncated, let us assume that \cite{Tawfik:2025kae,Tawfik:2024itt,Tawfik:2023hdi,Tawfik:2023ugm}. 
\bea
\tilde{g}_{\mu \nu}  &=& \left(\phi^2(p)+2\frac{\kappa}{(p_0^0)^2} K^2\right) 
 \left[1 + \frac{\dot{p}_0^{\mu} \dot{p}_0^{\nu} }{{\mathscr F}^2} \left(1+2\beta p_0 ^{\rho} p_{0 \rho}\right) \right] g_{\mu\nu} \equiv C(x,p) g_{\mu\nu}. \label{eq:gmunuQ3}
\eea 
Incorporating the proposed quantum-mechanical revisions, the conformal rescaling $C(x,p)$ of $g_{\mu\nu}$ exhibits a clear dependence on the eight-dimensional (phase space) spacetime. In this context, $\bar{m}$ refers to the mass of the free particle normalized to the Planck mass, denoted as $m_{\ell}$. The quantity ${\mathscr F}$ represents the discovery of the maximal proper force, which corresponds to the maximum proper acceleration experienced by the free-falling quantum particle in curved spacetime. Both quantities are influenced by the proposed quantization and emerge additional curvatures. The free particle's motion is governed by ${\mathscr F}$, which is impacted by the additional curvatures introduced through the proposed quantization process. This connection is established through the association with the maximal proper acceleration ${\mathscr A}=2\pi(c^7/G\hbar)^{1/2}$ discovered by Caianiello \cite{Caianiello:1981jq,caianiello1984maximal,brandt1989maximal}. The appearance of ${\mathscr F}$ and ${\mathscr A}$ as novel physical constants is a consequence of the quantum geometric approach to curved spacetime \cite{Rosen:1962mpv,Caianiello:1989wm,Caianiello:1989pu}.

The function $\phi(p)=1+\beta p_0^{\rho} p_{0\rho}$ represents a significant discovery 
\begin{itemize}
\item By incorporating the function $\phi(p)$, the Heisenberg uncertainty principle (HUP) is extended to RGUP. This expansion enables the examination of the effects of relativistic energy and finite gravitational fields on quantum mechanics \cite{Tawfik:2021ekh}.
\item Through the quantum mechanical revision of the Finsler metric tensor, $\phi(p)$ introduces a generalization of the fundamental tensor, thereby expanding the framework of GR \cite{Tawfik:2023rrm,Tawfik:2023kxq}. 
\item By maintaining the unique curvature properties of the Randers metric, $\phi(p)$ effectively combines gravity and electromagnetism into a unified framework \cite{Cheng2012},
\bea
\phi(p) &=& 1 + \beta p_0^{\rho} p_{0 \rho} = 1 + \frac{\kappa}{(p_0^0)^2} K^2.
\eea
\end{itemize} 

\subsection{Derivation of Quantum mechanical Revision of Finsler--Hamilton Metric}
\label{sec:FnslMtr}

In the context of Klein metric $K$, the coordinates and momenta of a free-falling quantum particle with a finite positive mass $m$ are represented by auxiliary four-vectors, namely $x_0^{\alpha}$ and $p_0^{\beta}$, respectively \cite{Klein1910},
\bea
K^2(x_0^{\alpha}, p_0^{\beta}) &=& \frac{\left\| p_0^{\beta}\right\|^2 - \left\|x_0^{\alpha}\right\|^2 \left\| p_0^{\beta}\right\|^2 + \left\langle x_0^{\alpha} \cdot p_0^{\beta}\right\rangle^2}{1-\| x_0^{\alpha}\|^2}. \label{eq:Klein1}
\eea
The standard Euclidean norm and inner product in $\R^n$ are denoted by $\|\cdot\|$ and $\langle \cdot\rangle$, respectively. In order to maintain $0$-homogeneity of $\phi(p)$, the RGUP method can be directly utilized for the $1$-homogeneous function $F(x_0^{\alpha}, p_0^{\beta})$ with respect to $p_0^{\beta}$.
\bea
K^2(x_0^{\alpha}, \phi(p) p_0^{\beta}) &=& \frac{\left\| \phi p_0^{\beta}\right\|^2 - \left\| x_0^{\alpha}\right\|^2 \left\| \phi p_0^{\beta}\right\|^2 + \left\langle x_0^{\alpha} \cdot \phi p_0^{\beta}\right\rangle^2}{1-\left\| x_0^{\alpha}\right\|^2}  = \phi^2(p) K^2(x_0^{\alpha}, p_0^{\beta}). \label{eq:Klein2}
\eea
The quantum mechanical revision of eight-dimensional Finsler metric is obtained by considering the Hessian matrix of the squared $K(x_0^{\alpha}, \phi(p) p_0^{\beta})$
\bea
\tilde{g}_{\alpha \beta}(x,p)&=&\left(\phi^2(p) + \frac{2 \kappa}{(p_0^0)^2} K^2\right) g_{\alpha \beta}(x,p) \nn \\
&+& \frac{2 \kappa}{(p_0^0)^2} \left[4 \phi(p) K^2 \ell_{\alpha} \ell^{\sigma} g_{\sigma \alpha}(x) - 4 \phi(p) K^3 \ell^{\sigma} \left[\delta_{0 \beta} g_{\sigma \alpha}(x,p) + \delta_{0 \alpha} g_{\sigma \beta}(x,p)\right] \right. \nn \\
&&\left. \hspace*{9mm} + K^4[2+\phi(p)] \delta_{0 \alpha} \delta_0^{\sigma} g_{\sigma \beta}(x,p) \right], \label{eq:FnslrG1} 
\eea
where
\bea
\ell_{\gamma} &=& \frac{p_0^{\gamma}}{F} + \frac{\langle x_0, p_0\rangle}{\left(1-|x_0|^2\right) K} x_0^{\gamma}, \\
\phi_{\mu} &=& \frac{2\kappa K}{(p_0^0)^3} (p_0^0\ell_{\mu}-F\delta_{0\mu}), \\
\phi_{\mu\nu} &=& \frac{2\kappa }{(p_0^0)^2} g_{\mu\nu}(x,p)-\frac{4\kappa K}{(p_0^0)^3}(\ell_\nu \delta_{0\mu} + \ell_\mu \delta_{0\nu}) + \frac{6\kappa K^2 }{(p_0^0)^4} \delta_{0\nu}\delta_{0\mu}.
\eea

Because $\phi(p)$ depends solely on $p_0$ rather than $x_0$, it follows that the second line in Eq. (\ref{eq:FnslrG1}) should not vanish, categorically. Due to the intricate nature of the geometric structure and the presence of additional curvatures as expressed in the second line, coupled with the absence of a theoretical framework to fully address the entirety of expression (\ref{eq:FnslrG1}), one is compelled to accept this unavoidable approximation. Only through this bold approximation can we gain insight into the implications of the proposed quantization on the Riemann manifold. However, once it becomes feasible to incorporate all terms of Eq. (\ref{eq:FnslrG1}), it becomes necessary to reexamine the resulting quantum mechanical revision of four-dimensional Riemann metric, as described in Eq. (\ref{eq:gmunuQ2}).

The section that follows examines how the proposed geometric quantization influences the affine connection.

\section{Geometric Quantization and Affine Connections}
\label{sec:GGA}

In accordance with section \ref{sec:tildegmunuApp}, the suggestion was made to utilize the quantum mechanically revisited metric tensor at a given point $x$ as a comprehensive representation of the curved spacetime in Riemann geometry, encompassing all relevant details about the geometry of the spacetime \cite{Tawfik:2025kae,Tawfik:2024itt,Tawfik:2023ugm,Tawfik:2023hdi,Tawfik:2023rrm,Tawfik:2023kxq}
\bea
\tilde{g}_{\alpha\beta} &=& C(x,p)\, g_{\alpha\beta}. \label{eq:tildegalphabeta1}
\eea
The determination of the exact expression for $C(x,p)$ continues to pose a significant mathematical challenge. As an alternative, an approximate formulation has been put forward in the references \cite{Tawfik:2025kae,Tawfik:2024itt,Tawfik:2023ugm,Tawfik:2023hdi,Tawfik:2023rrm,Tawfik:2023kxq}
\bea
C(x,p) &=&\left(\phi^2(p)+\frac{2 \kappa}{(p_0^0)^2} K^2\right) 
 \left[1 + \frac{\dot{p}_0^{\mu} \dot{p}_0^{\nu}}{{\mathscr F}^2} \left(1+2\beta p_0 ^{\rho} p_{0 \rho}\right)  \right].
\label{eq:Ctildegalphabeta1}
\eea
It is noteworthy that $C(x,p)$ can be averaged. Various techniques for averaging exist. For further exploration of this subject, we invite readers to review our publications \cite{Tawfik:2023hdi,Tawfik:2023ugm,Tawfik:2023rrm,Tawfik:2023kxq}. While this approach may lead to a bias towards a specific instance within the continuous spectrum ranges of the quantum operators forming $C(x,p)$, the truncated $C(x,p)$ continues to exhibit its important quantum-mechanical features.

The assumption of linearity in Eq. (\ref{eq:tildegalphabeta1}) implies that the relationship $-c^2 d\tau^{2} = ds^{2}=g_{\alpha\beta} dx^{\alpha} dx^{\beta}$ remains applicable to the quantum mechanically revisited metric tensor $\tilde{g}_{\alpha\beta}$. By appropriately parameterizing, the proper time can be represented - in natural units - as $- d\tau^{2} = ds^{2}$
\bea    
\tilde{\tau}_{ab} &=& \int_{0}^{1}\sqrt{-\tilde{g}_{\alpha\beta}(x) \frac{dx^{\alpha}}{d\sigma}\frac{dx^{\beta}}{d\sigma}} = \int_{0}^{1}L\left(\frac{dx^{\alpha}}{d\sigma},x^{\alpha}\right)d\sigma.
\eea
By employing variational methods, similar to those used in classical dynamics, one can derive the Euler--Lagrange equations 
\bea
-\frac{d}{d\sigma} \frac{\partial L}{\partial (dx^{\gamma}/d\sigma)}+\frac{\partial L}{\partial x^{\gamma}}&=&0, \label{eq:L}
\eea
where $\frac{\partial L}{\partial x^{\gamma}}$ and $\frac{d}{d\sigma}\frac{\partial L}{\partial (d x^{\gamma}/d\sigma)}$ are given as follows:
\bea
\frac{\partial L}{\partial x^{\gamma}} &=& \frac{-L}{2}\left\{
C(x,p) \frac{\partial g_{\alpha \beta}}{\partial x^{\gamma}}\frac{d x^{\alpha}}{d\tau}\frac{d x^{\beta}}{d\tau}  + g_{\alpha \beta}\frac{2\kappa}{(p_{0}^0)^{2}}\left[
1+\frac{\tilde{m}^{2}}{{\mathscr F}^{2}}\left(1+2 \beta p_{0}^{\rho} p_{0}^{\rho}\right)\right]
F^{2}_{,\gamma} \frac{d x^{\alpha}}{d\tau}\frac{d x^{\beta}}{d\tau}\right\}, \nn \label{eq:L1} \hspace*{7mm} \\
\frac{d}{d\sigma}\frac{\partial L}{\partial (d x^{\gamma}/d\sigma)} &=& -L\left[\tilde{g}_{\alpha\gamma}\frac{d^{2} x^{\alpha}}{d\tau^{2}}+\frac{1}{2}\left(\frac{\partial \tilde{g}_{\alpha\gamma} }{\partial x^{\beta}}+\frac{\partial \tilde{g}_{\gamma\beta} }{\partial x^{\alpha}}\right)\frac{d x^{\alpha}}{d\tau}\frac{d x^{\beta}}{d\tau}\right]. \nn \label{eq:L2}
\eea

The quantum mechanically revisited geodesic equations are derived by substituting equations (\ref{eq:L1}) and (\ref{eq:L2}) into equation (\ref{eq:L})
\bea
\frac{d^{2}x^{\alpha}}{d\tau^{2}} + \tilde{\Gamma}^{\alpha}_{\delta \beta} \frac{dx^{\delta}}{d\tau}\frac{dx^{\beta}}{d\tau} + \frac{g_{\delta \beta}}{2 \tilde{g}_{\alpha \gamma}} K^{2}_{,\gamma} C(x,p) \frac{dx^{\delta}}{d\tau}\frac{dx^{\beta}}{d\tau} &=& 0. \label{eq:C1}
\eea
The geodesic equations obtained from the conventional metric tensor, $g_{\alpha\beta}$, 
\bea
\frac{d^{2}x^{\alpha}}{d\tau^{2}}+\Gamma^{\alpha}_{\delta \beta} \frac{dx^{\delta}}{d\tau}\frac{dx^{\beta}}{d\tau}&=& 0, \label{eq:ClsGeoEq0}
\eea
and those obtained from $\tilde{g}_{\alpha\beta}$, as expressed in Eq. (\ref{eq:C1}), exhibit a remarkable difference. This refers to the fulfillment of the geodesic equations by $\tilde{g}_{\alpha\beta}$. Detailed explanations will be presented in Appendix \ref{sec:App1}. The fulfillment of the geodesic equations by $\tilde{g}_{\alpha\beta}$ clearly establishes that EFE are also satisfied. Additionally, ref. \cite{Tawfik:2023ugm} presents a comprehensive analysis of the EFE on a three-sphere with $\tilde{g}_{\alpha\beta}$. It was concluded that the EFE are optimally satisfied when both versions of the metric tensor are considered.
Such a comprehensive difference encompasses not only the entire third term in Eq. (\ref{eq:C1}), but also the additional quantum-mechanical ingredients and curvatures that arise from the quantum mechanically revisited affine connections $\tilde{\Gamma}^{\alpha}_{\delta \beta}$ \cite{Tawfik:2023ugm,Tawfik:2023hdi,Tawfik:2023rrm,Tawfik:2023kxq}.
\bea
\tilde{\Gamma}^{\alpha}_{\delta \beta}= 
\Gamma^{\alpha}_{\delta \beta}+\frac{K^{2}_{, \gamma}}{2 C(x,p)}\left(\delta^{\alpha}_{\beta}+\delta^{\alpha}_{\delta}-g^{\alpha \gamma} g_{\delta \beta}\right). \label{eq:C}
\eea
Consequently, Equation (\ref{eq:C1}) can be reformulated as,
\bea
\frac{d^{2}x^{\alpha}}{d\tau^{2}}+\Gamma^{\alpha}_{\delta \beta} \frac{dx^{\delta}}{d\tau}\frac{dx^{\beta}}{d\tau} + \frac{K^{2}_{, \gamma}}{2 C(x,p)}\left[\frac{g_{\delta \beta}}{\tilde{g}_{\alpha \gamma}} C^2(x,p) \frac{dx^{\delta}}{d\tau}\frac{dx^{\beta}}{d\tau} + \delta^{\alpha}_{\beta}+\delta^{\alpha}_{\delta}-g^{\alpha \gamma} g_{\delta \beta}\right] &=& 0. \hspace*{7mm} \label{eq:C2}
\eea
The third term represents the total contributions that have arisen from the proposed geometric quantization approach.

Affine connections and their modifications due to the proposed geometric quantization are fundamental in determining the timelike geodesic congruence, as outlined in section \ref{sec:egsSol1}. We first introduce the Einstein--Gilbert--Straus metric in section \ref{sec:EGS1}.

\subsection{Einstein--Gilbert--Straus Metric}
\label{sec:EGS1}

The integration of Schwarzschild and FLRW metrics enables the embedding of the Schwarzschild metric within a Universe that is undergoing expansion. This methodology provides a realistic representation of the Universe. In the 1940s, Einstein and Straus proposed the Swiss cheese model, which aimed to model the Universe as clusters of cosmic matter that are unevenly distributed alongside empty regions or voids \cite{RevModPhys.17.120}. Gilbert successfully formulated the Schwarzschild metric in an expanding Universe by employing the conventional metric tensor $g_{\alpha\beta}$ \cite{Gilbert1956,Stewart:1990ngn}. In contrast, the FLRW solution describes the expansion or contraction of the homogeneous and isotropic Universe uniformly, regardless of the observation location. Alternatively, the EGS metric provides a theoretical framework for the inclusion of nonhomogeneity in the context of the Universe
\bea
ds^{2}=\left(1-\frac{2M}{r}\right)dt^{2}-\left(1-\frac{2M}{r}\right)^{-1} a(t) dr^{2}-a(t)r^{2} d\Omega^2, \label{eq:Glbrt1}
\eea
where $d\Omega^2=d\theta^{2}+\sin^{2}{\theta}d\phi^{2}(p)$ represents the standard metric on the surface of a two-sphere with a constant radius $r$. The scale factor $a(t)$ determines the rate at which the size of the geometric structure of the Universe can change relative to its original size. Apparently, $t$ represents the cosmic time. The substance that fills the cosmic background influences the underlying scale factor \cite{Tawfik:2011sh,Tawfik:2010pm,Tawfik:2010bm}. In the current calculations, a matter-dominated cosmic background is assumed, meaning that $a(t)$ is proportional to $t^{2/3}$. The mass $M$ corresponds to a spherical mass around which EFE are solved. In this scenario, the mass is represented by the symbol $M$, which serves to distinguish it from the $m$ that refers to the free-falling quantum particle. At $a(t)=1$ in Eq. (\ref{eq:Glbrt1}), the spherically symmetric vacuum Schwarzschild solution can be retrieved, straightforwardly. This particular solution describes the gravitational field in the vicinity of a spherical distribution of mass $M$. The spatial and temporal distribution of the substance can be interpreted as the total cosmic substance.  

As the equation (\ref{eq:Glbrt1}) indicates a dependence on both $r$ and $t$ in describing the evolution of the Universe, it follows that the cosmic substance $M$ should also vary accordingly. We can denote this variation as $M(r,t)$ to incorporate it into the EGS metric. Accordingly, one has to account for the spatial and temporal evolution of our Universe. Another consequence of $M(r,t)$ in the EGS metric is the necessity of a finite cosmological constant $\Lambda$, that causes expansion and evolution of the Universe \cite{Diab:2020jcl,Tawfik:2011sh,Tawfik:2011mw}. The value that would be attributed to $\Lambda$ does not impose any further limitations on our model. Therefore, with $M(r,t)$ just nonvanishing $\Lambda$ improves our modeling of EGS.

To simplify the analysis we propose that: i) the cosmic substance follows the barotropic equation of state $p=-\rho$, where $p$ represents pressure and $\rho$ represents energy density, and ii) the Einstein-Straus vacuole forms exact spheres \cite{PhysRev.136.B571,osti_4325043},
\bea
M(r,t) &=& \frac{4}{3} \pi f(t) r^3. \label{eq:Mrt1}
\eea 
This outlines a basic framework for understanding the distribution of cosmic substance in the context of the Einstein--Straus vacuole.
We put forth a third assumption that the arbitrary function $f(t)$ provides a depiction of the temporary evolution of the cosmic substance. To accomplish this, it is essential to proportionally adjust the radius $r$ in relation to the passage of time $t$. One of the authors (AT) proposed a model in which the scaling function of the matter distribution is represented as \cite{NasserTawfik:2024afw}
\bea
f(t) &=& \frac{1}{\mu^2} \tanh\left(\frac{t}{\mu}\right). \label{eq:Mrt2}
\eea 
where the parameter $\mu$ is a variable selected without particular limitations; however, it is essential for maintaining the correct mass dimension in Eq. \eqref{eq:Mrt1}.  As discussed, the concept of an expanding Universe is closely tied to the presence of a finite $\Lambda$ and the scale factor $a(t)$. Consequently, the static metric defined in Eq. (\ref{eq:Glbrt1}) can be transformed into a dynamic representation
\bea
ds^{2}=\left(1-\frac{2M(r,t)}{r}-\frac{1}{3} \Lambda r^2\right)dt^{2} - a(t) \left(1-\frac{2M(r,t)}{r}-\frac{1}{3} \Lambda r^2\right)^{-1} dr^{2} - a(t)r^{2} d\Omega^2, \label{eq:Glbrt2}
\eea

\subsection{Timelike Geodesic Congruence and Expansion Evolution with $g_{\alpha \beta}$}
\label{sec:egsSol1}

In the case of comoving coordinates, where the system of coordinates moves with the cosmic substance at each point, all components of the four-velocity disappear except for the first component. Specifically, for the spherical coordinates $x^{\alpha}=(t,r,\theta,\phi)$, it follows that $u^t=u^{\theta}=u^{\phi}=0$ \cite{PhysRev.136.B571}. With variable separation, the first geodesic equation can then be expressed as
\bea
\frac{d u^t(r,t)}{u^t(r,t)} &=& - \frac{\partial M(r,t)}{\partial t} \frac{3 dt}{6M(r,t)+\Lambda r^3-3r}. \label{eq:1stGeoConvg}
\eea
Then, we get
\bea
u^t(r,t) &=& \left(r^2\left[\Lambda + 8 \pi \tanh(t)\right] - 3\right)^{-1/2}.
\eea
It is evident that the real value of $u^t(r,t)$, for any given $r$, is conditioned to
\bea
t & > & \tanh^{-1}\left(\frac{1}{8 \pi}\left[\frac{3}{r^2} - \Lambda\right]\right). \label{eq:Cond1}
\eea
The derivation of the geodesic congruence expansion $\Theta(r,t)$ involves examining the velocity fields and affine connections, resulting from the expression $\Theta(r,t) = u^{\alpha}(r,t)_{;\alpha}$, with $u^{\alpha}(r,t)$ representing the velocity vector in the spacetime manifold
\bea
\Theta(r,t) &=& u^{\alpha}(r,t)_{,\alpha} + u^{\sigma}(r,t) \Gamma^{\alpha}_{\sigma \alpha}. \label{eq:clTheta01}
\eea
By substitution the derivations for $u^{\alpha}_{,\alpha}$ and $u^{\sigma} \Gamma^{\alpha}_{\sigma \alpha}$ into Eq. \eqref{eq:clTheta01}, we obtain
\bea
\Theta(r,t) &=& \frac{\left(u^t(r,t)\right)^{3}}{2 r a(t)} \left\{
\frac{3 r \dot{a}(t)}{\left(u^t(r,t)\right)^{2}} - 6 a(t) \frac{\partial M(r,t)}{\partial t} \right\}, \label{eq:clTheta1}
\eea 
where $\dot{a}(t)$ denotes the time derivative of the scale factor.

Now, let us derive the equation that describes the evolution of the geodesic congruence expansion, specifically $\partial \Theta(r,t)/\partial r$. By applying the ordinary chain rule, which is motivated by physical interdependence and mathematical principle, we obtain that 
\bea
\frac{d\Theta(r,t)}{d\tau} &=& \frac{\partial \Theta(r,t)}{\partial r} \frac{\partial r}{\partial t} \frac{\partial t}{\partial \tau}. \label{eq:apprntHrz1}
\eea 
As mentioned earlier, the third derivative can be identified with $u^t(r,t)$. On the other hand, the second derivative can be approximated, for instance, at the proper horizon \cite{Kaloper:2010ec}. We make the assumption that the apparent horizon, which is the boundary where at least one null-geodesic congruence undergoes a change in its focusing properties, serves as the boundary condition for determining $\partial r/\partial t$. This assumption ensures that the corresponding null constraint is satisfied, specifically at $ds^2=0$, as shown in Eq. (\ref{eq:Glbrt1}). The condition of vanishing $ds^2$ guarantees the vanishing of $d\Omega^2$. Then, the second derivative in equation \eqref{eq:apprntHrz1} can be expressed as follows:
\bea
\frac{dr}{dt} &=& \pm \frac{1}{\sqrt{a(t)}} \left(1-2\frac{M(r,t)}{r}-\frac{1}{3} \Lambda r^2\right), \label{eq:drdt1}
\eea
where $\pm$ indicate the direction of the rays, with the plus sign denoting outgoing rays and the minus sign denoting ingoing rays. Hence, the evolution equation for $\Theta(r,t)$ can be determined.
\bea
\frac{d\Theta(r,t)}{d\tau} &=& \frac{\left(u^t(r,t)\right)^{4}}{2 r^2 \left(a(t)\right)^{3/2}} \left\{
6 a(t) \frac{\partial M(r,t)}{\partial t} + \frac{2 r a(t)}{\left(u^t(r,t)\right)^{2}} \frac{\partial^2 M(r,t)}{\partial r \partial t} \right. \nn \\
&&\left. \hspace*{18mm} + r^2 (\Lambda + 8 \pi \tanh(t)) \left[\frac{r \dot{a}(t)}{\left(u^t(r,t)\right)^{2}} - 8 a(t) \frac{\partial M(r,t)}{\partial t} \right] 
\right\}. \label{eq:dthetadrdt1}
\eea
This analytic expression shall be evaluated and depicted in Fig. \ref{fig:1}.

The subsequent section outlines the derivation of timelike geodesic congruence and expansion evolution, employing the quantum mechanically revisited metric tensor of GR $\tilde{g}_{\alpha \beta}$.

\subsection{Timelike Geodesic Congruence and Expansion Evolution with $\tilde{g}_{\alpha \beta}$}
\label{sec:egsSol2}

The proposed quantum mechanically revisited metric tensor, Eq. (\ref{eq:tildegalphabeta1}), allows for the derivation of $\tilde{u}^t(r,t)$ from the relevant geodesic equation
\bea
\frac{d \tilde{u}^t(r,t)}{\tilde{u}^t(r,t)} &=& - \frac{\partial M(r,t)}{\partial t} \frac{3 dt}{6 M(r,t) +\Lambda r^3 -3r} - \frac{K^2_{, \gamma}}{2 C(x,p)} dt - \frac{1}{2} K^2_{, \gamma} dt, \nn \\
&=& \frac{d u^t(r,t)}{u^t(r,t)} - \left(\frac{K^2_{, \gamma}}{2 C(x,p)} + \frac{1}{2} K^2_{, \gamma}\right) dt,
\eea
where $C(x,p)$ represents the conformal coefficient that connects the quantum mechanically revisited and conventional metric tensor, while $K^2_{, \gamma}$ stands for the comma derivative of the Klein metric, which is one of the simplest metrics in Finsler space. The Finsler structure is henceforth identified with the Klein metric.  Solving this initial first-order ordinary differential equation leads to
\bea
\tilde{u}^t(r,t) &=& \left(r^2\left[\Lambda + 8 \pi  t \tanh(t) \right]-3\right)^{-1/2} \exp\left(-\frac{1+C(x,p)}{2 C(x,p)} t K^2_{, \gamma}\right), \nn \\
&=& u^t(r,t) \exp\left(-\frac{1+C(x,p)}{2 C(x,p)} t K^2_{, \gamma}\right) \label{eq:tldu1}
\eea 
In comparison to Eq. \eqref{eq:clTheta01}, the quantized geodesic congruence expansion $\tilde{\Theta}(r,t)$ can be represented as  
\bea
\tilde{\Theta}(r,t) &=& \tilde{u}^{\alpha}(r,t)_{,\alpha} + \tilde{u}^{\sigma}(r,t) \tilde{\Gamma}^{\alpha}_{\sigma \alpha}. \label{eq:clTheta}
\eea 

Analogous to Eq. \eqref{eq:apprntHrz1}, the evolution equation for the quantum mechanically revisited geodesic congruence expansion can be derived as
\bea
\frac{d\tilde{\Theta}(r,t)}{d\tau} &=& \frac{\partial \tilde{\Theta}(r,t)}{\partial r} \frac{\partial r}{\partial t} \frac{\partial t}{\partial \tau}, \label{eq:apprntHrz2}
\eea 
where in this case $\partial t/\partial \tau$ represents $\tilde{u}^1(r,t)$, Eq. (\ref{eq:tldu1}), and $\partial r/\partial t$ could also be estimated, at the proper horizon, Eq. (\ref{eq:drdt1}) \cite{Kaloper:2010ec}. Consequently, the expansion evolution is presented as
\bea
\frac{d \tilde{\Theta}(r,t)}{d \tau} &=& - \frac{r \left(u^t(r,t)\right)^{2}}{6 C(x,p) a^{3/2}(t)} \exp\left(- 2\frac{1+C(x,p)}{2 C(x,p)} t K^2_{, \gamma} \right) \nn \\
& & \left\{\left[a(t) \left(C(x,p)-3\right) K^2_{, \gamma} - 3 C(x,p) \dot{a}(t)\right] \frac{\Lambda + 8 \pi t \tanh(t)}{\left(u^t(r,t)\right)^{2} \exp\left(-2\frac{1+C(x,p)}{2 C(x,p)} t K^2_{, \gamma} \right)} \right. \nn \\
& & \hspace{0mm} \left. + 8 \pi C(x,p) a(t) \left[9+\frac{\left(u^t(r,t)\right)^{-2}}{\exp\left(-2\frac{1+C(x,p)}{2 C(x,p)} t K^2_{, \gamma} \right)}\right] \left[t \sech^2(t) + \tanh(t)\right] 
\right\}.
\eea
This analytical expression will be subjected to numerical analysis in \ref{sec:dsct}. The section that follows provides an in-depth exploration of insights related to quantum mechanical revisions that should be undertaken for the quantization of general relativity and also addresses some averaging techniques, especially the mean-field approximation.

\subsection{Aspects of Quantum Mechanical Revision and Mean-Field Approximation}
\label{sec:qMFA}

The ultimate aim of this research is to reconcile the principles of QM with those of GR. Our geometric quantization approach concerning GR necessitates two principal generalizations. The first generalization is intended to incorporate gravitational aspects into the fundamental framework of quantum mechanics. The second generalization explores the notion of quantum geometry, which entails the inclusion of additional curvatures, the dependence on acceleration or gravitational force of the free-falling quantum particle in curved spacetime. This framework enables the simulation of the proposed quantization in Finsler--Hamilton and then translated into Riemann spacetime. To procedure to express the resulting metric tensor on a Riemann manifold, we apply the ratio of the lengths of any two collinear vectors, which do not depend on the underlying metric tensor. In this regard, we assume that the measure of line elements on both manifolds are {\it ''exact''} and therefore yield deterministic outcomes. Thus, the ultimate quantization is evidently dependent on the quantization of the line element. Although the current version of quantization tackles the identified challenges and integrates comprehensive discretization, it is still only partially aligned with quantum mechanical revisions.
 
To achieve quantization of the line element, it is essential to integrate a probability distribution with the metric tensor and the $1$-form $d x^{\mu}$. As an alternative, the suggestion of noncommutation relations may be necessary. In this context, it is important to note that neither $g_{\mu \nu}$ nor $dx^{\mu}$, including its generalized form, exhibits a clear noncommutation under translation \cite{Martinetti:2009tr}. Conversely, the establishment of a noncommutative differential calculus \cite{Dubois-Violette:1999sns,Madore:2000aq} alongside a noncommutative metric tensor \cite{Ulhoa:2015rza} may facilitate the integration of these approaches into the formulation of a noncommutative measure for the line element \cite{FitzGerald:2005uvh}. Moreover, the revised relativistic kinematics, characterized geometrically by Finsler geometry, provides insights into the possible vacuum state of quantum gravity at low energies \cite{Pfeifer:2021tas}. The adoption of a length element in Finsler geometry in conjunction with RGUP discretization seems to lead to a quantization of spacetime. A thorough examination of this topic is warranted in future research.

For the sake of simplicity, the quantum mechanical revisited metric tensor is given on expectation value, $\langle g\rangle^{\mu \nu}$, which is a mean-field approximation. This means that the $\langle g\rangle^{\mu \nu}$ represents the metric tensor whose indexes are assessed based on the expected value. This implies that $\langle g\rangle^{\mu \nu}$ is built by operators rather than classical variables \cite{maddox1975schur}. 

The Quantum operators are hypothesized to characterize spacetime states as a quantum system \cite{Gine:2021nhu}
\bea
\langle \hat{g}(\hat{x})\rangle^{\mu \nu} &\equiv & \langle g\rangle^{\mu \nu}(\hat{x}),
\eea
where $\hat{x}$ are the quantum operators which act on the underlying spacetime. It is crucial to highlight that our formalism depends on $\langle \hat{g}(\hat{x})\rangle^{\mu \nu}$. Furthermore, a formalism for expectation values was suggested in recent literature \cite{Gine:2021nhu}, which is independent of the signature of the metric tensor. The quantum operator computes the averaged metric tensor for a given $\mu$. The other index $\nu$ contains information on the operator's dimensions. Furthermore, the possibility of the existence of a discrete spectrum for the metric and the probability distribution or average values of the metric could be considered elsewhere. 

However, it is important to note that despite the introduction of a bias towards a specific instance within the continuous spectrum range of quantum operators, the process of averaging does not completely obscure their quantum-mechanical nature. This is true. But such an averaging procedure seems to ensure that coherence in the Klein metric and the conformal coefficient, for instance, is likely to be maintained. It is worth mentioning that this particular averaging process bears similarities to the well-known {\it ''measurement problem''} in QM. In simpler terms, it is an inevitable consequence of the collapse of quantum characteristics that occurs during the measurement process. Looking ahead, once the mathematical challenge of deriving the quantum fundamental tensor, as presented in section \ref{sec:tildegmunuApp}, is resolved, addressing the continuous spectrum range of the quantum operators should also be tackled.

To conclude, it is essential to note that despite the ongoing mathematical and physical difficulties these factors present, we are obliged to accept their compromises, as they seem to aid in the pursuit of resolving the long-standing dilemma of integrating the principles of GR and QM. This is particularly significant considering the substantial truncation required in the derivation of the quantum fundamental tensor, as discussed in section \ref{sec:tildegmunuApp}, and the quantization of line element measures in Finsler--Hamilton and Riemann geometries, as elaborated in this section. Further investigation into all these aspects may be conducted in other studies.

The forthcoming section will undertake a comprehensive numerical analyses that explore the cosmological and astrophysical implications of our geometric quantization approach, focusing on both spatial and initial singularities.

\section{Numerical Results and Discussion}
\label{sec:dsct}

The present manuscript is specifically intended to address the cosmological and astrophysical implications of our geometric quantization approach. We discuss a systematic examination of the time-like (initial) singularity, as well as the investigation of space-like (spacial) singularity. Therefore, we tackled one of the most realistic solutions of EFE that simulate the evolution and expansion of our Universe.

\begin{figure}[htb!] 
\includegraphics[angle=-90,width=0.96\textwidth]{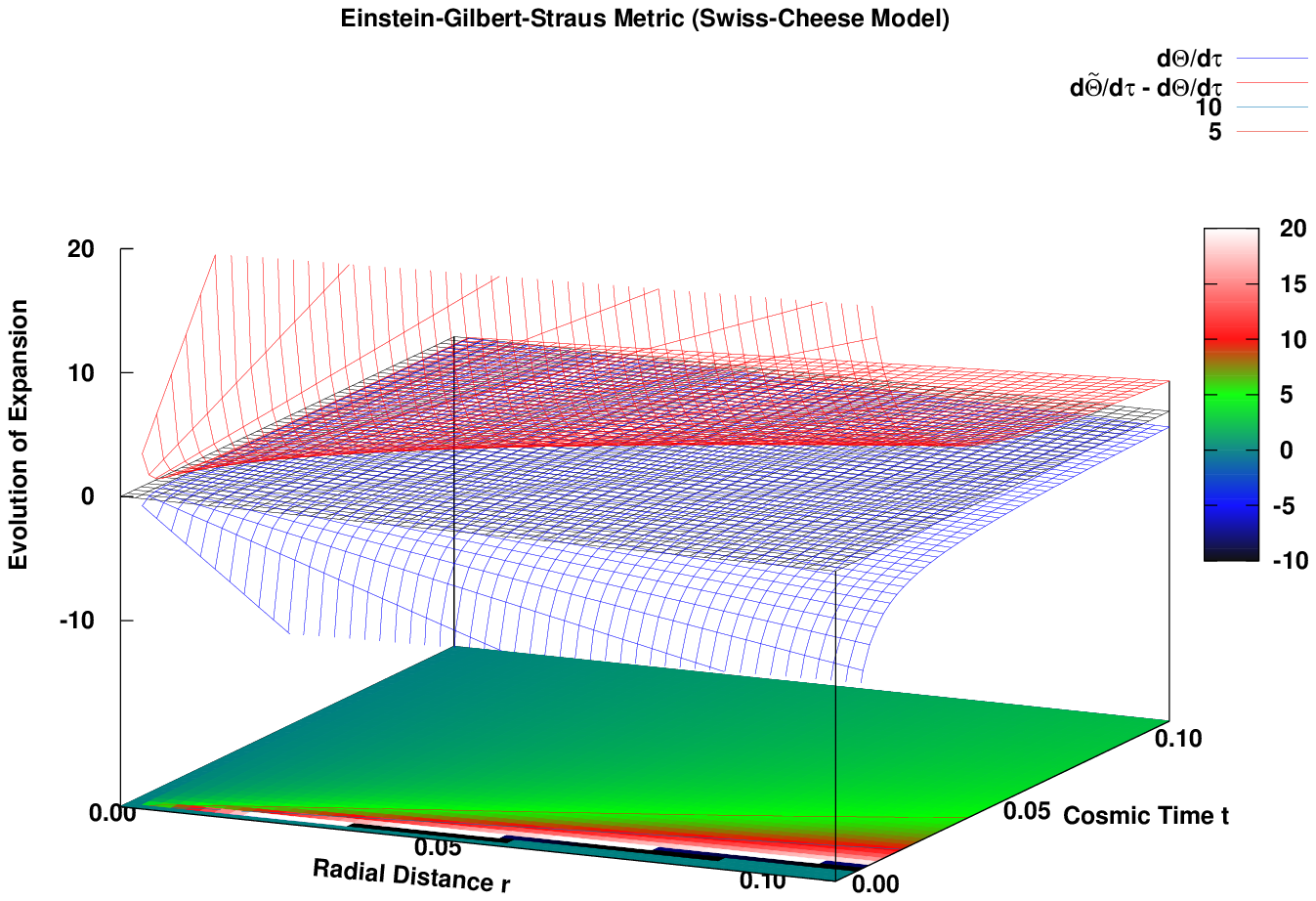} 
\caption{The evolution of the expansion of the timelike geodesic congruence, $d\Theta/d\tau$, represented as a function of the radial distance $r$ and cosmic time $t$ (blue lattice), whose analytical expression is given in Eq. \eqref{eq:dthetadrdt1}. The difference $d\tilde{\Theta}(r,t)/d \tau - d\Theta(r,t)/d \tau$, which indicates the pure quantum contribution (red lattice) is confronted with. A black lattice presents vanishing expansion evolution. This depiction serves to indicate which metric tensor is connected to the emergence or removal of space and initial singularity of GR. \label{fig:1}}
\end{figure}

In addition to the analytic evolution of the expansion $\Theta(r,t)$ and its quantized version $\tilde{\Theta}(r,t)$ outlined in the previous section, this section will present the numerical analysis of $d\Theta(r,t)/d \tau$ and $d\tilde{\Theta}(r,t)/d \tau$. In Fig. \ref{fig:1}, the comparison is made between the exclusive contribution from the conventional metric tensor and that from the proposed quantization. This comparison allows for the evaluation of the relative impacts of the proposed geometric quantization. The quantity $d\tilde{\Theta}(r,t)/d \tau$ includes $d\Theta(r,t)/d \tau$ obtained under vanishing quantization, in addition to the pure quantum mechanically inspired contributions. As a result, a comparison between $d\Theta(r,t)/d \tau$ (blue lattice) and $d\tilde{\Theta}(r,t)/d \tau - d\Theta(r,t)/d \tau$ (red lattice) is conducted to emphasize the contributions added by the proposed geometric quantization in assessing or removing the space and initial singularity from GR. Moreover, we depict the total disappearance of expansion evolution (shown as a black lattice) to aid in understanding the specific fundamental tensor that results in either positive or negative evolution of expansion. This is a critical assessment of the presence or absence of spacetime singularity.

For the purpose of this numerical analysis, dummy values are suggested for $M$, $\mu$, and $\Lambda$, for instance. Additionally, the quantum operators $K^2_{, \alpha}$ and $C(x,p)$ are averaged, specifically by mean-field approximation. Consequently, this process is likely to have an impact on the qualitative numerical analysis of these operators.

In Fig \ref{fig:1}, we present a comparison between the results derived from the conventional metric tensor (blue tensor) and those obtained using the quantum mechanically revisited metric tensor (red lattice). It is evident that the evolution of the expansion $\Theta(r,t)$ with respect to $\tau$ is significant, particularly for small values of radial distance $r$ and cosmic time $t$ (blue lattice). As $r$ and $t$ decrease, the resulting $d\Theta(r,t)/d\tau$ becomes increasingly negative. According to the focusing theorem, the subsequent evolution of expansion leads to the inference that there exists a presence of rapidly converging congruence and continuous geodesics. This outcome lends support to the proposition that the Swiss cheese model, the EGS solution of EFE, appears to uphold the preservation of space and the initial singularity in curved spacetime. This suggests that the corresponding expansion likely experiences a swift transition between outgoing (diverging) and ongoing geodesic (converging congruence). Essentially, the curvature of spacetime governed by cosmic substance is presumed to oscillate between open and closed configurations. Consequently, we conclude that the expansion promptly oscillates between reduced divergence and rapid convergence. The strong focusing associated with this scenario indicates a significant gravitational attraction.

The evolution equation of the timelike geodesic congruence for the quantized fundamental tensor, as denoted by Eq. (\ref{eq:tildegalphabeta1}), consistently shows a positive sign across all regions. Specifically, the comparison between the pure quantum contribution, $d\tilde{\Theta}(r,t)/d\tau - d\Theta(r,t)/d\tau$ (red lattice), and $d\Theta(r,t)/d\tau$ (blue lattice), highlights the effects of the proposed geometric quantization. As both $r$ and $t$ decrease, the timelike geodesic congruence undergoes an increased level of positivity in its evolution. This feature is specifically linked to an outgoing geodesic, underscoring the divergence within the congruence and indicating the absence of spatial and initial singularity.

\section{Conclusions and Outlook}
\label{sec:cncl}

Analytical and numerical analyses are employed in this study to investigate the spacetime curvature as a function of radial distance and cosmic time, assuming that the cosmic background consists of lumps of substance nonhomogeneously distributed within voids and holes. It is postulated that the magnitude of the substance in the lumps is equivalent to that excavated to form the holes. The Swiss cheese model, initially proposed by Einstein and Straus and further refined by Gilbert, is utilized to realistically represent our Universe. The EGS metric extends the conventional cosmological model by incorporating the nonhomogeneous aspects of the expanding and evolving Universe. The complexity inherent in this solution of EFE seems to limit its prominence among theorists and cosmologists, especially when contrasted with other less complicated metrics.

By employing the conventional and quantum mechanically revisited metric tensor, the timelike geodesic congruence within the EGS metric is employed to characterize the spacetime curvature and investigate the intrinsicality of space and initial singularity. The utilization of this metric enables the simultaneous examination of the evolution of world lines with respect to both radial distance $r$ and cosmic time $t$. In the present investigation, it is postulated that the cosmic substance contained within the nonhomogeneously voids is characterized by exact spheres, whose radii undergo variations due to the expansion of the Universe. 

The application of a quantum geometric approach and a kinematic theory of free-falling quantum particle on an extended eight-dimensional manifold leads to quantum-mechanically induced revisions on the corresponding metric, resulting in a quantum mechanically revisited metric tensor expressed as a conformal transformation of the conventional one. The eight-dimensional manifold, along with the metric derived from it, has been fully discretized and is at least partially quantized. Through the mean-field approximation, the conformal coefficient can be effectively averaged, facilitating calculations related to the evolution of the timelike geodesic congruence. The outcomes obtained from the conventional fundamental tensor consistently show negativity, suggesting the presence of space and initial singularity. In contrast, the quantum mechanically revisited metric tensor demonstrates a positive sign, particularly at small values of both $r$ and $t$, indicating the absence of space and initial singularity. A negative evolution denotes a reduction in the previously expanding state, whereas a positive evolution signifies a continuation of the expansion. Thus we  conclude that the proposed geometric quantization seems to provide a viable solution to the spacetime curvature resulting from both spatial and initial singularities of GR. Moreover, this quantization process gives rise to additional curvatures, which act as sources of gravity, i.e., quantum gravity.

As a future outlook, efforts should be focused on two critical aspects: investigating the prospective quantum features of the extra curvature and addressing the mathematical complexities that arise in the process of deriving the quantum mechanically revisited metric tensor through applying bold truncation and averaging procedures. Further future studies can target reinforcing the proposed geometric quantization. In detail, we plan to examine the quantum aspects of the quantized metric tensor. We have addressed several components of the proposed geometric quantization, all of which will be explored in greater depth in forthcoming research.


\appendix

\section{Geodesic Equations and Quantum Mechanically Revisited Metric Tensor}
\label{sec:App1}

We start the appropriate normalization condition,
\bea
C(x,p)\, g_{\mu\nu}\, u^{\mu}\, u^{\nu} &=& -1.
\eea
Then, we apply the quantum mechanically revisited metric tensor, 
\bea 
\widetilde{g}_{\mu\nu} &=& C(x,p)\, g_{\mu\nu}. \label{eq:G2} 
\eea
This functionality is required to assist in the extraction of geodesic equations that have been revisited from a quantum mechanical perspective. The covariant differentiation results in
\bea
C(x,p) g_{\mu\nu} u^{\mu} u^{\nu} \widetilde{\nabla}_{\gamma}  + C(x,p) g_{\mu\nu} u^{\mu} \widetilde{\nabla}_{\gamma} u^{\nu}
 & + & \nn \\
  C(x,p) g_{\mu\nu} u^{\nu} \widetilde{\nabla}_{\gamma}  u^{\mu}  + C(x,p)  u^{\mu} u^{\nu} \widetilde{\nabla}_{\gamma} g_{\mu\nu} &=& 0, \label{eq:nabla1}
\eea
where $\widetilde{\nabla}_{\gamma} u^{\mu} = u^{\mu}_{, {\gamma}} + u^{\sigma} \widetilde{\Gamma}^{\nu}_{\sigma k}$.   
By eliminating the last terms in this expression, we find that
\bea
C(x,p) g_{\mu\nu} u^{\mu} u^{\nu} \widetilde{\nabla}_{\gamma} + 2 C(x,p) g_{\mu\nu} u^{\mu} \widetilde{\nabla}_{\gamma} u^{\nu} &=&0, \nn \\
2 C(x,p) g_{\mu\nu} u^{\mu} \widetilde{\nabla}_{\gamma} u^{\nu} &=& - g_{\gamma \kappa} u^{{\gamma}} u^{\kappa} \widetilde{\nabla}_{\gamma} C(x,p), \nn \\
2 u^{\mu} \widetilde{\nabla}_{\gamma} u^{\nu} &=& - \frac{g_{\gamma \kappa}}{\widetilde{g}_{\mu \nu}} u^{{\gamma}} u^{\kappa} \widetilde{\nabla}_{\gamma} C(x,p) \nn
\eea
Thus, the geodesic equations can be represented as 
\bea
\frac{d^2 x^{\mu}}{d\tau^2} + \widetilde{\Gamma}^{\mu}_{{\gamma} \kappa} \frac{d x^{\gamma}}{d\tau} \frac{d x^{\kappa}}{d\tau} + \frac{g_{\gamma \kappa}}{\widetilde{g}_{\mu \nu}} C(x,p)_{, {\gamma}} \frac{d x^{\gamma}}{d\tau} \frac{d x^{\kappa}}{d\tau} &=& 0. \label{eq:GeoCond}
\eea
It is evident that Eq. (\ref{eq:GeoCond}) is evidently equivalent to Eq. (\ref{eq:C1}). Hence, we can conclude that the geodesic equations can be extracted from the proposed generalized normalization condition, thereby showcasing their adherence to the principles of conservation laws.




\section*{Acknowledgment}

AT acknowledges the generous support by the Egyptian Center for Theoretical Physics (ECTP) and Future University in Egypt (FUE)!

\section*{Conflicts of Interest}

The authors declare that there are no conflicts of interest regarding the publication of this published article. The authors declare that the present script is in compliance with ethical standards regarding its content!

\section*{Dataset Availability}

All data generated or analyzed during this study are included in this published article. The data used to support the findings of this study are included within the published article and properly cited! All of the material is owned by the authors.

\section*{Author contributions}

The responsibility for proposing the conception of the present study lies with AT, who also undertook the tasks of designing and managing the research, interpreting the results, deriving the expressions, drawing the figures, and preparing the manuscript. AA and AP contributed to the writing and proofreading of the manuscript. The final version of the manuscript was unanimously approved by all authors.

\section*{Funding}

The authors declare that this research received no specific grants from any funding agency in the public, commercial, or not-for-profit sectors.


\section*{Competing interests}

The authors confirm that there are no relevant financial or non-financial competing interests to report.

\bibliographystyle{unsrtnat}
\bibliography{TDThesis-ListOfReferences-Motion}

\begin{thebibliography}{49}
\providecommand{\natexlab}[1]{#1}
\providecommand{\url}[1]{\texttt{#1}}
\expandafter\ifx\csname urlstyle\endcsname\relax
  \providecommand{\doi}[1]{doi: #1}\else
  \providecommand{\doi}{doi: \begingroup \urlstyle{rm}\Url}\fi

\bibitem[Tawfik and Abou El~Dahab(2017)]{Tawfik:2017ngn}
A.~Tawfik and E.~Abou El~Dahab.
\newblock {FLRW Cosmology with Horava-Lifshitz Gravity: Impacts of Equations of
  State}.
\newblock \emph{Int. J. Theor. Phys.}, 56\penalty0 (7):\penalty0 2122--2139,
  2017.
\newblock \doi{https://doi.org/10.1007/s10773-017-3355-1}.

\bibitem[McVittie(1933)]{McVittie:1933zz}
G.~C. McVittie.
\newblock {The mass-particle in an expanding universe}.
\newblock \emph{Mon. Not. Roy. Astron. Soc.}, 93:\penalty0 325--339, 1933.
\newblock \doi{10.1093/mnras/93.5.325}.

\bibitem[Tolman(1934)]{Tolman1934}
Richard~C. Tolman.
\newblock Effect of inhomogeneity on cosmological models.
\newblock \emph{Proc Natl. Acad. Sci. USA}, 20:\penalty0 169--176, 1934.
\newblock \doi{10.1073/pnas.20.3.169}.

\bibitem[Einstein and Straus(1945)]{RevModPhys.17.120}
Albert Einstein and Ernst~G. Straus.
\newblock The influence of the expansion of space on the gravitation fields
  surrounding the individual stars.
\newblock \emph{Rev. Mod. Phys.}, 17:\penalty0 120--124, Apr 1945.
\newblock \doi{https://doi.org/10.1103/RevModPhys.17.120}.
\newblock URL \url{https://link.aps.org/doi/10.1103/RevModPhys.17.120}.

\bibitem[Bondi(1947)]{Bondi:1947fta}
H.~Bondi.
\newblock {Spherically symmetrical models in general relativity}.
\newblock \emph{Mon. Not. Roy. Astron. Soc.}, 107:\penalty0 410--425, 1947.
\newblock \doi{10.1093/mnras/107.5-6.410}.

\bibitem[Gilbert(1956)]{Gilbert1956}
C.~Gilbert.
\newblock {The Gravitational Field of a Star in the Expanding Universe}.
\newblock \emph{Monthly Notices of the Royal Astronomical Society},
  116\penalty0 (6):\penalty0 678--683, 12 1956.
\newblock ISSN 0035-8711.
\newblock \doi{10.1093/mnras/116.6.678}.
\newblock URL \url{https://doi.org/10.1093/mnras/116.6.678}.

\bibitem[Stephani and Stewart(1990)]{Stewart:1990ngn}
H.~Stephani and J.~Stewart.
\newblock { Allgemeine Relativitätstheorie}.
\newblock \emph{Cambridge University Press}, 1990.

\bibitem[Nasser~Tawfik et~al.(2024)Nasser~Tawfik, Pasqua, Waqas, Alshehri, and
  Kr~Haldar]{NasserTawfik:2024afw}
Abdel Nasser~Tawfik, Antonio Pasqua, Muhammad Waqas, Azzah~A. Alshehri, and
  Prabir Kr~Haldar.
\newblock {Quantum geometric perspective on the origin of quantum-conditioned
  curvatures}.
\newblock \emph{Class. Quant. Grav.}, 41\penalty0 (19):\penalty0 195018, 2024.
\newblock \doi{10.1088/1361-6382/ad7451}.

\bibitem[Tawfik and Dabash(2023{\natexlab{a}})]{Tawfik:2023onh}
A.~Tawfik and T.~F. Dabash.
\newblock {Timelike geodesic congruence in the simplest solutions of general
  relativity with quantum-improved metric tensor}.
\newblock \emph{Int. J. Mod. Phys. D}, 32\penalty0 (15):\penalty0 2350097,
  2023{\natexlab{a}}.
\newblock \doi{10.1142/S0218271823500979}.

\bibitem[Tawfik and Dabash(2023{\natexlab{b}})]{Tawfik:2023ugm}
Abdel~Nasser Tawfik and Tahia~F. Dabash.
\newblock {Born reciprocity and discretized Finsler structure: An approach to
  quantize GR curvature tensors on three-sphere}.
\newblock \emph{Int. J. Mod. Phys. D}, 32\penalty0 (10):\penalty0 2350068,
  2023{\natexlab{b}}.
\newblock \doi{10.1142/S0218271823500682}.

\bibitem[Tawfik and Dabash(2023{\natexlab{c}})]{Tawfik:2023hdi}
Abdel~Nasser Tawfik and Tahia~F. Dabash.
\newblock {Born reciprocity and relativistic generalized uncertainty principle
  in Finsler structure: Fundamental tensor in discretized curved spacetime}.
\newblock \emph{Int. J. Mod. Phys. D}, 32\penalty0 (09):\penalty0 2350060,
  2023{\natexlab{c}}.
\newblock \doi{10.1142/S0218271823500608}.

\bibitem[Tawfik et~al.(2025)Tawfik, Dabash, Amer, and Shaker]{Tawfik:2025kae}
Abdel~Nasser Tawfik, Tahia~F. Dabash, Tarek~S. Amer, and Mohamed~O. Shaker.
\newblock {Einstein-Gilbert-Straus solution of Einstein field equations:
  Timelike geodesic congruence with conventional and quantized fundamental
  metric tensor}.
\newblock \emph{Nucl. Phys. B}, 1014:\penalty0 116866, 2025.
\newblock \doi{10.1016/j.nuclphysb.2025.116866}.

\bibitem[Tawfik et~al.(2024)Tawfik, Farouk, Tarabia, and Maher]{Tawfik:2024itt}
Abdel~Nasser Tawfik, Fady~T. Farouk, F.~Salah Tarabia, and Muhammad Maher.
\newblock {Quantum-induced revisiting space\textendash{}time curvature in
  relativistic regime}.
\newblock \emph{Int. J. Mod. Phys. A}, 39\penalty0 (35):\penalty0 2443016,
  2024.
\newblock \doi{10.1142/S0217751X24430164}.

\bibitem[Tawfik(2023{\natexlab{a}})]{Tawfik:2023kxq}
Abdel~Nasser Tawfik.
\newblock {On possible quantization of the fundamental tensor in the
  relativistic regime}.
\newblock \emph{Astron. Nachr.}, 344\penalty0 (1-2):\penalty0 e220072,
  2023{\natexlab{a}}.
\newblock \doi{10.1002/asna.20220072}.

\bibitem[Tawfik(2023{\natexlab{b}})]{Tawfik:2023rrm}
Abdel~Nasser Tawfik.
\newblock {On quantum-induced revisiting Einstein tensor in the relativistic
  regime}.
\newblock \emph{Astron. Nachr.}, 344\penalty0 (1-2):\penalty0 e220071,
  2023{\natexlab{b}}.
\newblock \doi{https://doi.org/10.1002/asna.20220071}.

\bibitem[Farouk et~al.(2023)Farouk, Tawfik, Tarabia, and Maher]{Farouk:2023hmi}
Fady~Tarek Farouk, Abdel~Nasser Tawfik, Fawzy~Salah Tarabia, and Muhammad
  Maher.
\newblock {On Possible Minimal Length Deformation of Metric Tensor, Levi-Civita
  Connection, and the Riemann Curvature Tensor}.
\newblock \emph{MDPI Physics}, 5\penalty0 (4):\penalty0 983--1002, 2023.
\newblock \doi{10.3390/physics5040064}.

\bibitem[Tawfik et~al.(2021)Tawfik, Farouk, Tarabia, and Maher]{Tawfik:2021ekh}
Abdel~Nasser Tawfik, Fady~T. Farouk, F.~Salah Tarabia, and Muhammad Maher.
\newblock {Minimal length discretization and properties of modified metric
  tensor and geodesics}.
\newblock In \emph{{16th Marcel Grossmann Meeting on~Recent Developments in
  Theoretical and Experimental General Relativity, Astrophysics and
  Relativistic Field Theories}}, 11 2021.
\newblock \doi{https://doi.org/10.1142/9789811269776_0339}.

\bibitem[Tawfik and Alshehri(2024)]{Tawfik:2024gxc}
Abdel~Nasser Tawfik and Azzah Alshehri.
\newblock {Relativistic generalized uncertainty principle for a test particle
  in four-dimensional spacetime}.
\newblock \emph{Mod. Phys. Lett. A}, 39\penalty0 (25n26):\penalty0 2450079,
  2024.
\newblock \doi{10.1142/S0217732324500792}.

\bibitem[Kolomytsev(1972)]{Kolomytsev:1972rf}
V.~I. Kolomytsev.
\newblock {On the problem of a family of nonparallel regge trajectories}.
\newblock \emph{Teor. Mat. Fiz.}, 12:\penalty0 40--47, 1972.
\newblock \doi{10.1007/BF01030039}.

\bibitem[Tsamparlis and Grammenos(1995)]{Tsamparlis:1995nq}
M.~Tsamparlis and Th. Grammenos.
\newblock {The Deviation equation for a general congruence in a general
  spacetime}.
\newblock \emph{Tensor (Japan)}, 56:\penalty0 27--30, 1995.

\bibitem[Hawking and Penrose(1970)]{Hawking:1970zqf}
S.~W. Hawking and R.~Penrose.
\newblock {The Singularities of gravitational collapse and cosmology}.
\newblock \emph{Proc. Roy. Soc. Lond. A}, 314:\penalty0 529--548, 1970.
\newblock \doi{10.1098/rspa.1970.0021}.

\bibitem[Landsman(2022)]{Landsman:2022hrn}
Klaas Landsman.
\newblock {Penrose\textquoteright{}s 1965 singularity theorem: from geodesic
  incompleteness to cosmic censorship}.
\newblock \emph{Gen. Rel. Grav.}, 54\penalty0 (10):\penalty0 115, 2022.
\newblock \doi{10.1007/s10714-022-02973-w}.

\bibitem[Borissova(2024)]{Borissova:2023kzq}
Johanna~N. Borissova.
\newblock {Suppression of spacetime singularities in quantum gravity}.
\newblock \emph{Class. Quant. Grav.}, 41\penalty0 (12):\penalty0 127002, 2024.
\newblock \doi{10.1088/1361-6382/ad46c0}.

\bibitem[DeWitt(1967)]{DeWitt:1967yk}
Bryce~S. DeWitt.
\newblock {Quantum Theory of Gravity. 1. The Canonical Theory}.
\newblock \emph{Phys. Rev.}, 160:\penalty0 1113--1148, 1967.
\newblock \doi{10.1103/PhysRev.160.1113}.

\bibitem[Nelson and Regge(1991)]{Nelson:1991an}
J.~E. Nelson and T.~Regge.
\newblock {(2+1) quantum gravity}.
\newblock \emph{Phys. Lett. B}, 272:\penalty0 213--216, 1991.
\newblock \doi{10.1016/0370-2693(91)91822-D}.

\bibitem[Caianiello(1981)]{Caianiello:1981jq}
E.~R. Caianiello.
\newblock {Is There a Maximal Acceleration?}
\newblock \emph{Lett. Nuovo Cim.}, 32:\penalty0 65, 1981.
\newblock \doi{https://doi.org/10.1007/BF02745135}.

\bibitem[Caianiello(1984)]{caianiello1984maximal}
E.~R. Caianiello.
\newblock Maximal acceleration as a consequence of heisenberg’s uncertainty
  relations.
\newblock \emph{Lettere al Nuovo Cimento (1971-1985)}, 41\penalty0
  (11):\penalty0 370--372, 1984.

\bibitem[Brandt(1989)]{brandt1989maximal}
Howard~E Brandt.
\newblock Maximal proper acceleration and the structure of spacetime.
\newblock \emph{Foundations of Physics Letters}, 2\penalty0 (1):\penalty0
  39--58, 1989.

\bibitem[Rosen(1962)]{Rosen:1962mpv}
Nathan Rosen.
\newblock {Quantum geometry}.
\newblock \emph{Annals Phys.}, 19\penalty0 (1):\penalty0 165--172, 1962.
\newblock \doi{https://doi.org/10.1016/0003-4916(62)90235-X}.

\bibitem[Caianiello et~al.(1990{\natexlab{a}})Caianiello, Feoli, Gasperini, and
  Scarpetta]{Caianiello:1989wm}
E.~R. Caianiello, A.~Feoli, M.~Gasperini, and G.~Scarpetta.
\newblock {Quantum Corrections to the Space-time Metric From Geometric Phase
  Space Quantization}.
\newblock \emph{Int. J. Theor. Phys.}, 29:\penalty0 131, 1990{\natexlab{a}}.
\newblock \doi{https://doi.org/10.1007/BF00671323}.

\bibitem[Caianiello et~al.(1990{\natexlab{b}})Caianiello, Gasperini, and
  Scarpetta]{Caianiello:1989pu}
E.~R. Caianiello, M.~Gasperini, and G.~Scarpetta.
\newblock {Phenomenological Consequences of a Geometric Model With Limited
  Proper Acceleration}.
\newblock \emph{Nuovo Cim. B}, 105:\penalty0 259, 1990{\natexlab{b}}.
\newblock \doi{https://doi.org/10.1007/BF02726101}.

\bibitem[Cheng and Shen(2012)]{Cheng2012}
Xinyue Cheng and Zhongmin Shen.
\newblock \emph{Randers Metrics with Special Riemann Curvature Properties},
  pages 77--89.
\newblock Springer Berlin Heidelberg, Berlin, Heidelberg, 2012.
\newblock \doi{https://doi.org/10.1007/978-3-642-24888-7_6}.

\bibitem[Klein(1910)]{Klein1910}
Felix Klein.
\newblock Über die geometrischen grundlagen der lorentzgruppe.
\newblock \emph{Jahresbericht der Deutschen Mathematiker-Vereinigung},
  19:\penalty0 533--552, 1910.
\newblock \doi{https://doi.org/10.1007/978-3-642-51960-4_31}.

\bibitem[Tawfik and Harko(2012)]{Tawfik:2011sh}
A.~Tawfik and T.~Harko.
\newblock {Quark-Hadron Phase Transitions in Viscous Early Universe}.
\newblock \emph{Phys. Rev. D}, 85:\penalty0 084032, 2012.
\newblock \doi{10.1103/PhysRevD.85.084032}.

\bibitem[Tawfik et~al.(2010)Tawfik, Wahba, Mansour, and Harko]{Tawfik:2010pm}
A.~Tawfik, M.~Wahba, H.~Mansour, and T.~Harko.
\newblock {Hubble Parameter in QCD Universe for finite Bulk Viscosity}.
\newblock \emph{Annalen Phys.}, 522:\penalty0 912--923, 2010.
\newblock \doi{10.1002/andp.201000103}.

\bibitem[Tawfik et~al.(2011)Tawfik, Wahba, Mansour, and Harko]{Tawfik:2010bm}
A.~Tawfik, M.~Wahba, H.~Mansour, and T.~Harko.
\newblock {Viscous Quark-Gluon Plasma in the Early Universe}.
\newblock \emph{Annalen Phys.}, 523:\penalty0 194--207, 2011.
\newblock \doi{10.1002/andp.201000052}.

\bibitem[Diab and Tawfik(2022)]{Diab:2020jcl}
Abdel~Magied Diab and Abdel~Nasser Tawfik.
\newblock {A Possible Solution of the Cosmological Constant Problem Based on
  GW170817 and Planck Observations with Minimal Length Uncertainty}.
\newblock \emph{Adv. High Energy Phys.}, 2022:\penalty0 9351511, 2022.
\newblock \doi{https://doi.org/10.1155/2022/9351511}.

\bibitem[Tawfik(2011)]{Tawfik:2011mw}
A.~Tawfik.
\newblock {The Hubble parameter in the early universe with viscous QCD matter
  and finite cosmological constant}.
\newblock \emph{Annalen Phys.}, 523:\penalty0 423--434, 2011.
\newblock \doi{10.1002/andp.201100038}.

\bibitem[Misner and Sharp(1964)]{PhysRev.136.B571}
Charles~W. Misner and David~H. Sharp.
\newblock Relativistic equations for adiabatic, spherically symmetric
  gravitational collapse.
\newblock \emph{Phys. Rev.}, 136:\penalty0 B571--B576, Oct 1964.
\newblock \doi{10.1103/PhysRev.136.B571}.
\newblock URL \url{https://link.aps.org/doi/10.1103/PhysRev.136.B571}.

\bibitem[Demianski(1973)]{osti_4325043}
M~Demianski.
\newblock Some new solutions of the einstein equations of astrophysical
  interest.
\newblock \emph{Acta Astron.}, 23\penalty0 (3):\penalty0 197--232, 1973.
\newblock URL \url{https://www.osti.gov/biblio/4325043}.

\bibitem[Kaloper et~al.(2010)Kaloper, Kleban, and Martin]{Kaloper:2010ec}
Nemanja Kaloper, Matthew Kleban, and Damien Martin.
\newblock {McVittie's Legacy: Black Holes in an Expanding Universe}.
\newblock \emph{Phys. Rev. D}, 81:\penalty0 104044, 2010.
\newblock \doi{10.1103/PhysRevD.81.104044}.

\bibitem[Martinetti(2009)]{Martinetti:2009tr}
Pierre Martinetti.
\newblock {Line element in quantum gravity: The Examples of DSR and
  noncommutative geometry}.
\newblock \emph{Int. J. Mod. Phys. A}, 24:\penalty0 2792--2801, 2009.
\newblock \doi{10.1142/S0217751X09046242}.

\bibitem[Dubois-Violette(2001)]{Dubois-Violette:1999sns}
Michel Dubois-Violette.
\newblock {Lectures on graded differential algebras and noncommutative
  geometry}.
\newblock \emph{Math. Phys. Stud.}, 23:\penalty0 245--306, 2001.
\newblock \doi{10.1007/978-94-010-0704-7_15}.

\bibitem[Madore(2000)]{Madore:2000aq}
J.~Madore.
\newblock \emph{{An introduction to noncommutative differential geometry and
  its physical applications}}, volume 257.
\newblock 2000.

\bibitem[Ulhoa et~al.(2015)Ulhoa, Santos, and Amorim]{Ulhoa:2015rza}
S.~C. Ulhoa, A.~F. Santos, and R.~G.~G. Amorim.
\newblock {On Non-Commutative Correction of the G\"odel-type Metric}.
\newblock \emph{Gen. Rel. Grav.}, 47\penalty0 (9):\penalty0 99, 2015.
\newblock \doi{10.1007/s10714-015-1944-y}.

\bibitem[FitzGerald(2005)]{FitzGerald:2005uvh}
P.~L. FitzGerald.
\newblock {The Superfield quantisation of a superparticle action with an
  extended line element}.
\newblock \emph{Int. J. Mod. Phys. A}, 20:\penalty0 2639--2655, 2005.
\newblock \doi{10.1142/S0217751X05022263}.

\bibitem[Pfeifer and Relancio(2022)]{Pfeifer:2021tas}
Christian Pfeifer and Jos\'e~Javier Relancio.
\newblock {Deformed relativistic kinematics on curved spacetime: a geometric
  approach}.
\newblock \emph{Eur. Phys. J. C}, 82\penalty0 (2):\penalty0 150, 2022.
\newblock \doi{10.1140/epjc/s10052-022-10066-w}.

\bibitem[Maddox(1975)]{maddox1975schur}
I.~J. Maddox.
\newblock Schur's theorem for operators.
\newblock \emph{Bulletin Greek Math. Society}, 16:\penalty0 18--21, 1975.

\bibitem[Gine and Luciano(2022)]{Gine:2021nhu}
Jaume Gine and Giuseppe~Gaetano Luciano.
\newblock {Gravitational effects on the Heisenberg Uncertainty Principle: A
  geometric approach}.
\newblock \emph{Results Phys.}, 38:\penalty0 105594, 2022.
\newblock \doi{10.1016/j.rinp.2022.105594}.

\end{thebibliography}

\end{document}